\documentclass[aps,prl,twocolumn,groupedaddress,longbibliography]{revtex4-2}  
\usepackage{graphicx}  
\usepackage{bm}        
\usepackage{amssymb}   
\usepackage{xcolor}   
\usepackage{amsmath}   
\usepackage{hyperref} 
\usepackage{fancyhdr}
\begin{document}

\title{Transition from traveling fronts to diffusion-limited growth in
expanding populations} 
\author{Louis Brezin}
\affiliation{Department of Physics and Graduate Program in Bioinformatics, Boston University, Boston, Massachusetts 02215, USA}
\author{Kyle J. Shaffer}
\affiliation{Gettysburg College, Gettysburg, Pennsylvania 17325, USA}
\author{Kirill S. Korolev}
\affiliation{Department of Physics, Graduate Program in Bioinformatics, and Biological Design Center, Boston University, Boston, Massachusetts 02215, USA}
\email{korolev@bu.edu}

\begin{abstract}
Reaction-diffusion equations describe various spatially extended processes that unfold as traveling fronts moving at constant velocity. We introduce and solve analytically a model that, besides such fronts, supports solutions advancing as the square root of time. These sublinear fronts preserve an invariant shape, with an effective diffusion constant that diverges at the transition to linear spreading. The model applies to dense cellular aggregates of nonmotile cells consuming a diffusible nutrient. The sublinear spread results from biomass redistribution slowing due to nutrient depletion, a phenomenon supported experimentally but often neglected. Our results provide a potential explanation for the linear rather than quadratic increase of colony area with time, which has been observed for many microbes.
\end{abstract}

\maketitle 
\vspace{0.5\baselineskip}
\begin{center}
{\scriptsize DOI: https://doi.org/10.1103/p3tt-gq73 \\
Copyright \textcopyright\ 2011 by American Physical Society. All rights reserved.}
\end{center}
\vspace{0.5\baselineskip}
When non-motile cells grow, they form dense aggregates such as healthy tissues, tumors, biofilms, microbial mats, and colonies. The growth dynamics of such aggregates influence diverse phenomena, including disease onset and progression, agricultural productivity, geochemical cycles, and the integrity of human-built infrastructure~\cite{korolev:perspective, hall:bacterial_biofilms, sasse:root_microbiome, mendes:rhizosphere, ciofu:antimicrobial_resistance, lyons:oxygen_rise, falkowski:microbial_engines, flemming:biofouling}. Consequently, understanding these dynamics has been a focus of extensive research, employing both detailed application-specific models and simpler phenomenological frameworks aimed at uncovering general principles of population growth~\cite{muller:Db_model, jacob:review, tokita:phase_diagram, ohgiwari:phase_diagram, dukovski:metabolic_ring, xavier:framework, golden:morphologies, anderson:cancer_model, farrell:mechanical_colonies, kannan:gradients, warren:establishment, black:capillary, beroz:verticalization, pokhrel:biophysical, martinez:morphological}.

Among these approaches, reaction-diffusion equations have emerged as the dominant modeling paradigm, because they effectively incorporate nutrient diffusion, cellular growth and motility, mechanical interactions, and other key processes. Theoretical predictions has been most thoroughly tested in the context of microbial colonies due to their accessibility for quantitative measurement and manipulation. In particular, reaction-diffusion models have successfully explained complex pattern formations~\cite{jacob:review, tokita:phase_diagram, ohgiwari:phase_diagram, dukovski:metabolic_ring, golden:morphologies} and---perhaps most notably---the observed nearly constant expansion velocity of microbial colonies~\cite{pirt:kinetic, wakita:expansion, pipe:colony_models, korolev:amnat}. This constant front velocity is a striking prediction resulting from the interplay of diffusive transport and exponential growth.

Recent theoretical work has focused on how various biophysical processes, especially mechanical interactions, influence expansion velocity~\cite{farrell:mechanical_colonies, kannan:gradients, warren:establishment, black:capillary, beroz:verticalization, pokhrel:biophysical, martinez:morphological}. However, an increasing number of experiments suggest that the commonly assumed linear growth is not universal. In particular, many organisms under diverse growth conditions exhibit sublinear, power-law growth with an exponent close to one-half~\cite{kishony:colony_survey,pipe:colony_models,balazsi:yeast,dervaux:growth_form}. Here, we demonstrate that these experimental observations can be reconciled within the standard reaction-diffusion framework by incorporating the experimentally motivated dependence of biomass redistribution on nutrient concentration---a factor largely overlooked in previous models.


Although there are a great number of reaction-diffusion models of colony growth, they typically fall into one of three classes. The first class includes various generalizations of the Fisher-Kolmogorov-Petrovsky-Piskunov~(FKPP) equation~\cite{fisher:wave, kolmogorov:wave,murray:book}:

\begin{equation}
\frac{\partial b}{\partial t} = D_s\bm{\nabla}^2 b + r b\left(1-\frac{b}{K}\right).
\label{fisher_model}
\end{equation}

\noindent Here, the growth rate of biomass~$b$ is approximated by the standard logistic curve, which consists of exponential growth at low~$b$ and saturation at carrying capacity~$K$. The value of~$K$ is set by the initial nutrient concentration, which is not modeled explicitly. The motility is assumed to follow a random-walk-like pattern, with the effective diffusion constant given by~$D_s$. This classic equation was the first model of reaction-diffusion waves in population biology and motivated numerous subsequent studies in various fields~\cite{murray:book,saarloos:review}. It predicts invariant traveling fronts moving with velocity~$v=2\sqrt{D_sr}$ and an exponentially decreasing population density ahead of the wave. These predictions have been confirmed in many experimental and observational studies~\cite{giometto:tetrahymena, bosch:fkpp_ecology, gandhi:wave, wakita:expansion, hastings:synthesis, murray:book}, but only with motile organisms, e.g., bacteria swimming in very thin agar. In dense microbial colonies, the outward motion of cells is not diffusive, and population density abruptly drops to zero instead of showing a more gradual exponential decrease~\cite{pipe:colony_models, muller:Db_model, farrell:mechanical_colonies, kannan:gradients, warren:establishment, black:capillary, beroz:verticalization, pokhrel:biophysical, martinez:morphological}.  

To capture the sharp drop of the biomass at the front, density dependence was introduced in the diffusion term of the FKPP equation~\cite{murray:book,saarloos:review}:

\begin{equation}
\frac{\partial b}{\partial t} = D_p\bm{\nabla}\cdot(b\bm{\nabla} b) + r b\left(1-\frac{b}{K}\right),
\label{pressure_model}
\end{equation} 

\noindent where the new parameter~$D_p$ quantifies the emergent cooperative motility of the cells and could depend on many factors such as the agar concentration, surfactant production, and cell rigidity. Phenomenologically, the nonlinear diffusion could be explained by collective motion due to the repeated rearrangements of cells within the colony as they push against each other. Alternatively, the nonlinear diffusion can be derived from a hydrodynamic model that involves mechanical compression due to growth, friction with the substrate, and the flow of the biomass in response to mechanical forces~(see the Supplemental Material~\cite{si}). The front velocity in this model equals~$\sqrt{D_pr/2}$, and the population density vanishes linearly near the colony edge~\cite{kawasaki:exact_dispersal, petrovski:exact, birzu:dispersal, saarloos:review}; power law decay is also possible for slightly different models~\cite{si}.

Although Eq.~\eqref{pressure_model} recapitulated the growth of circular colonies reasonably well, it could not reproduce two essential aspects of colony growth. First, colonies stop growing well before reaching the edge of the Petri dish, and, second, colonies exhibit non-circular~(rough or branched) morphologies at low nutrient and high agar concentrations~\cite{jacob:review,tokita:phase_diagram, ohgiwari:phase_diagram,balazsi:yeast}. Both of these observations can be explained by nutrient limitation~\cite{jacob:review,tokita:phase_diagram, ohgiwari:phase_diagram,muller:Db_model}, which is introduced in the third class of models:

\begin{figure}
    \includegraphics{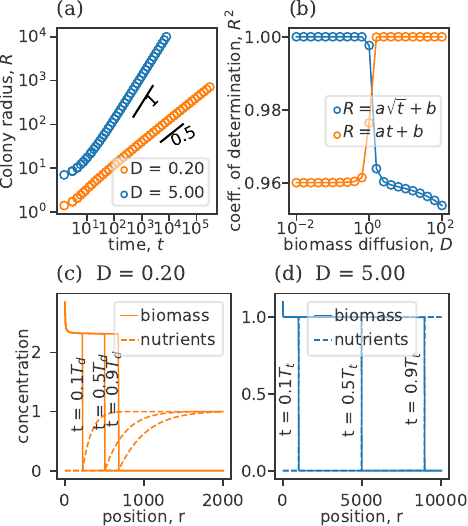}
\caption{(color online). Transition between traveling fronts and
diffusion-limited growth. (a)~Equation~\protect{\eqref{linear_model}}
supports two types of solutions with the colony radius $R$ 
increasing either linearly with time or as a square root of time.
(b)~The coefficient of determination~($R^2$) indicates that~$R\propto
\sqrt{t}$ for~$D<D_c$, and $R\propto t$ for~$D>D_c$. Examples of biomass and
nutrient profiles in the radial direction, at 10\%, 50\%, and 90\% of the time preceding
nutrient depletion at the end of the simulation box are shown in~(c) for~$D<D_c$ and in~(d) for~$D>D_c$.}
\label{fig:transition}
\end{figure}

\begin{equation}
\begin{aligned}
& \frac{\partial b}{\partial t} = D_p\bm{\nabla}\cdot(b\bm{\nabla} b) + \gamma bn,\\
& \frac{\partial n}{\partial t} = D_n\bm{\nabla}^2 n - \gamma bn .\\
\end{aligned}
\label{mvs_model}
\end{equation} 

\noindent Here,~$n$ is the concentration of the growth-limiting nutrient,~$D_n$ is its diffusion constant, and~$\gamma$ is the nutrient consumption rate. For simplicity, we assume that the biomass is measured in units such that one unit of nutrient produces one unit of biomass. We also neglect the metabolic cost of maintenance and the saturation of the nutrient uptake at high~$n$. These and other complications can be included, but they lead to similar dynamics; see the Supplemental Material~\cite{si}. 

Typically, small molecules diffuse much faster than the biomass. As a result, the nutrients are depleted in a region of about~$D_n/v$ ahead of the front, and the nutrient concentration within the colony is quite low. In fact, it decreases with~$D_n$, and the expansion velocity scales as~$D_p/\sqrt{D_n}$ in contrast to the other two models discussed above~\cite{muller:Db_model}.

Despite the great success of Eq.~\eqref{mvs_model} in capturing many
properties of microbial colonies, it only describes traveling fronts
moving with a constant velocity just like~Eqs.~(\ref{fisher_model})
and~(\ref{pressure_model})~\cite{jacob:review,tokita:phase_diagram,ohgiwari:phase_diagram,muller:Db_model};
see also Fig.~S2. While the radius of certain microbial colonies indeed grows linearly with time, it is the area that increases linearly in other experiments with seemingly
similar organisms and growth conditions~\cite{kishony:colony_survey,pipe:colony_models,balazsi:yeast,dervaux:growth_form,wang:morphologies}. 

The most extensive evidence for sublinear growth comes from the high-throughput screen of microbial growth patterns by Ernebjerg and Kishony~\cite{kishony:colony_survey} who measured how the radius of the colony increased with time for nearly five hundred colonies from soil isolates. The majority of the growth patterns were well described by a sublinear power law with exponent values clustered around~$0.5$. That is, the area of the colony rather than its diameter increased linearly with time for the majority of the colonies. The sublinear growth of the radius is often attributed to nutrient exhaustion, drying of the agar surface, or other experimental artifacts. While such mechanism should certainly be explored, it is unlikely that they explain these and other observations of a clearly linear increase of the area with time. In Ernebjerg and Kishony's experiments, colonies had very variable lag times and achieved variable final sizes, which means that they exhibited power law growth despite different stages of agar drying and varying degrees of nutrient depletion. 

The robustness against experimental artifacts is further confirmed by the studies of \textit{Bacillus subtilis} and \textit{Saccharomyces cerevisiae,}, two model organism whose growth has been carefully characterized by different experimental groups~\cite{balazsi:yeast,dervaux:growth_form,wang:morphologies}. Their results invariably show a very clean power law that starts after about ten hours of growth and continues for several days until either the experiment is terminated or the growth stops due to nutrient depletion.

To understand the origin of this common, but rarely characterized growth
pattern, we carefully examined the assumptions
behind~Eq.~\eqref{mvs_model}. Among all of the simplifications in this
model, the most questionable assumption is the functional form of the
biomass motility because it has not been carefully quantified. In fact,
existing experimental data strongly suggests that the rate of biomass
redistribution depends, perhaps indirectly, on the nutrient
concentration. The main evidence comes from experiments with two
identical strains labeled by different fluorescent
markers~\cite{hallatschek:sectors,korolev:amnat,korolev:rmp}. As the
colonies expand, demographic fluctuations lead to local fixation of one
of the strains and the establishment of monoclonal sectors. The
boundaries between these sectors are dynamic at the colony edge, where
cells are actively growing and the nutrient concentration is high, but
they are frozen in the colony bulk. Thus, biomass redistribution
requires active growth, which in turn requires nutrients. Further
supporting evidence comes from competition experiments with strains that
grow at different rates~\cite{korolev:sectors}. 
In this study, it was
found that the differences in expansion velocities are proportional to
the differences in the growth rates in liquid culture. This can be
reconciled with the predictions of
Eqs.~\eqref{fisher_model},~\eqref{pressure_model}, and~\eqref{mvs_model}
only when the biomass motility depends on the growth rate; otherwise,
the expected dependence is~$v\propto\sqrt{r}$ and~$v\propto\sqrt{\gamma
n}$ respectively, which does not match the experimental data.

Based on these observations, we modified~Eq.~\eqref{mvs_model} to capture the link between motility and growth as follows

\begin{equation}
\begin{aligned}
& \frac{\partial b}{\partial t} = D_b\bm{\nabla}\cdot(bn\bm{\nabla} b) + \gamma bn,\\
& \frac{\partial n}{\partial t} = D_n\bm{\nabla}^2 n - \gamma bn .\\
\end{aligned}
\label{linear_model}
\end{equation} 

\noindent Note that the effective diffusion coefficient has the same dependence on~$b$ and~$n$ as the growth term; in other words, we could say that the motility rate is proportional to the net growth rate. In the Supplemental Material, we discuss alternative formulations of this model and provide a derivation based on the balance of mechanical forces within the colony~\cite{si}. The only difference between this derivation and that of~Eq.~\eqref{pressure_model} is that we account not only for cell compression, but also for active stresses generated by colony growth.


To simplify the analysis, we nondimensionalize our model by measuring~$b$ and~$n$ in the units of the initial nutrient concentration~$n_0$, time in units of~$1/(\gamma n_0)$, and spatial positions in the units of~$\sqrt{D_n/(\gamma n_0)}$. This transformation sets~$n_0$ and all the coefficients in Eq.~\eqref{linear_model} to unity except for the biomass diffusivity, which becomes


\begin{equation}
D = \frac{D_bn^2_0}{D_n}.
\label{D}
\end{equation}

\noindent In the following, we use the nondimensionalized formulation without a change in notation. Note that the dimensional velocities are given by the nondimenionalized velocity~$v$ times the ``nutrient velocity'' given by~$v_n=\sqrt{D_n\gamma n_0}$. 

Using numerical simulations~\cite{si}, we examined the expansion of the biomass in $d$~spatial dimensions. We primarily focus on the expansions in narrow channels~$d=1$ and on the surface of a Petri dish~$d=2$, but other values of~$d$ are discussed in the Supplemental Material~\cite{si}. To simplify the calculations, our analytical and numerical analysis is focused on radially symmetric solutions. Thus, for~$d>1$, we assume that either growth instabilities do not occur or they are suppressed by large surface tension or other factors. This is a reasonable assumption because experiments show that mutations that reduce cellular adhesion result in perfectly circular colonies without any signs of instabilities even at low nutrient and high agar concentrations~\cite{balazsi:yeast}, i.e., in the regime where many organisms produce rough colonies with finger-like protrusion or branches. In the following, we use the radial coordinate~$r$ when we are describing the results for~$d>1$ and the linear coordinate~$x$ for the results specific to~$d=1$.

Figure~\ref{fig:transition}) shows our key result: different growth regimes for high and low~$D$. For large values of~$D$, the radius of the colony increased linearly with time, and the solution behaved as a standard reaction-diffusion front. Below a critical value of biomass diffusivity~$D_c\approx1$, the nature of the solution changed. The radius of the colony increased only as~$t^{1/2}$. Despite this slower growth, the spatial profile of the biomass density remained invariant in the co-moving reference frame, similar to a regular traveling front.

\begin{figure}
\includegraphics{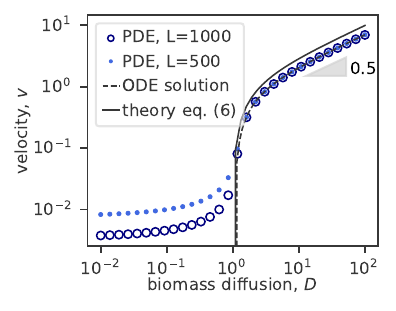}
\caption{(color online). Velocity of the traveling front solutions.
Circles show the velocities obtained by performing a linear fit
on~$R(t)$, determined by solving
Eq.~\protect{\eqref{linear_model}}~(PDE). These calculations were performed at two systems sizes. The disagreement between them indicates that a traveling front solution does not exist. The dashed line shows the results of the shooting method~(ODE), and the solid line is the analytical approximation given by Eq.~\protect{\eqref{v_analytic}}.}
\label{fig:velocity}
\end{figure}

First, we tested that Eq.~\eqref{linear_model} indeed admits solutions
of the form~$b(x-vt)$ and~$n(x-vt)$. Upon substituting the
traveling-front ansatz into the equation, we solved the resulting
ordinary differential equations both numerically using the shooting
method and analytically by making certain approximations; see the
Supplemental Material~\cite{si}. Both calculations confirmed that traveling front solutions exist only for~$D>D_c$ and showed the same behavior as the solutions of the time-dependent problem~(Fig.~\ref{fig:velocity}). 

The analytical solution for~$v(D)$ provides approximate, but very simple summary of our results:

\begin{equation}
v = \frac{D-1}{\sqrt{D}}.
\label{v_analytic}
\end{equation} 

\noindent In agreement with simulations,~$v\propto\sqrt{D}$ for
large~$D$, and the velocity vanishes at~$D_c=1$. The former scaling is
the same as for~Eq.~\eqref{pressure_model} because there is no nutrient
limitation in this regime. The large~$D$ behavior is, however, not
relevant for microbial colonies for which~$v<1$~(i.e., $v<v_n$ in
dimensional
units)~\cite{korolev:amnat,korolev:sectors,farrell:mechanical_colonies}.
Note that our results for~$D>D_c$ do not depend on the number of
    spatial dimensions because the traveling front solution emerges
    after a short transient when the radius of the colony is much larger
than the thickness of the growth front~\cite{si,murray:book}.

Second, we examined the solutions that exhibit the square-root growth.
Although both~$n$ and~$b$ moved forward as~$t^{1/2}$, only the biomass
profile remain invariant in time, just as for~$D>D_c$. The nutrient
profile was not a traveling wave. Instead, the nutrient concentration at
the edge of the colony decreased as~$t^{-1/2}$, and the region of
nutrient depletion ahead of the colony increases as~$t^{1/2}$; see
Figs.~S6 and S7. Hence, the length scale on which the nutrient varies becomes much larger than that on which the biomass varies, and we can simplify the problem by assuming that~$b(t,r)$ is a moving Heaviside step function:

\begin{equation}
b(t,r) = H\theta(r-r_e(t)),
\label{b_ansatz} 
\end{equation}

\noindent where~$r_e(t)$ is the position of the colony edge, and~$H$~is the biomass density within the colony, which reflects the height or thickness of an actual three-dimensional colony. The motion of the colony edge is then given by nutrient flux into the colony since nutrients are converted into biomass without loss in Eq.~\eqref{linear_model}:

\begin{equation}
H\frac{dr_e}{dt} = \left.\frac{\partial n}{\partial r}\right|_{r=r_e}.
\label{xe_equation} 
\end{equation}

To the leading order, the equation for~$n(t,r)$ becomes a simple diffusion equation with an absorbing~(Dirichlet) boundary condition at~$r=r_e(t)$. 

The simplified problem can be solved by the standard methods; see the Supplemental Material~\cite{si}, and the solution reads

\begin{equation}
r_e(t) = 2\varkappa\sqrt{t},
\label{xe_solution}
\end{equation}

\noindent where~$\varkappa$ is specified, in terms of~$H$, by the following equation

\begin{equation}
\begin{aligned}
H^{-1} = & 2\varkappa^d e^{\varkappa^2}\int_{\varkappa}^{+\infty}p^{1-d}e^{-p^2}dp \\
=&\left\{\begin{aligned} & \sqrt{\pi}\varkappa e^{\varkappa^2}\mathrm{erfc}(\varkappa) \;\; d=1\\ & \varkappa^2 e^{\varkappa^2}\mathrm{E_1}(\varkappa^2) \;\; d=2\end{aligned}\right.
\end{aligned}
\label{height}
\end{equation}

\noindent Here,~$\mathrm{erfc}(y)$ and~$\mathrm{E_1}(y)$ are the complementary error function and exponential integral respectively. The limiting behavior for small and large~$\varkappa$ is discussed in the Supplementary Material~\cite{si}. Briefly,~$H\to1$ from above for~$\varkappa\to+\infty$, and~$H\to+\infty$ when~$\varkappa\to0$. That is, colonies that expand more slowly are thicker. 

\begin{figure*}
    \includegraphics{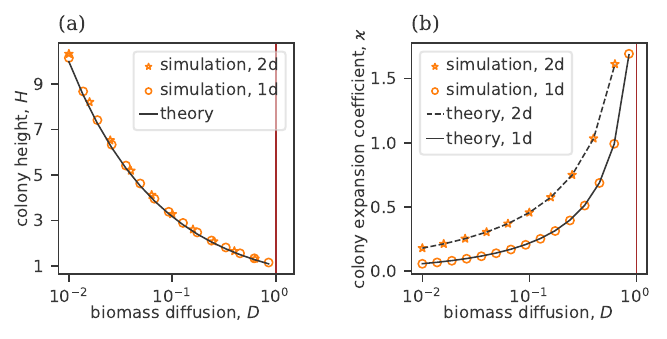}
\caption{Biomass diffusivity controls the height and expansion rate of
    microbial colonies in the regime of square-root growth, as
    demonstrated for simulations in one and two dimensions.
(a)~The colony height, defined as the maximal concentration near the
edge of the colony, diverges for~$D\to0$, as
expected for a stationary colony, and it approaches~$1$ for $D \to D_c$, as expected for
a traveling front solution. The solid line shows the analytical
prediction given by Eq.~\protect{\eqref{H_prediction}}.
(b)~The
rate of colony expansion, quantified by~$\varkappa=r_e/(2\sqrt{t})$,
increases with~$D$. It approaches zero for~$D=0$ and diverges at~$D\to
D_c$. Theoretical predictions based on
Eq.~\protect{\eqref{height}} are shown with a solid line for 1d and a
dashed line for 2d, using numerically obtained values of~$H$.}
\label{fig:dlg}
\end{figure*}

We can test~Eq.~\eqref{height} by obtaining~$H$ and~$\varkappa$ from
simulations for various values of~$D$, and then comparing the observed
values of~$H$ to the values of~$H$ predicted by~Eq.~\eqref{height}. This
comparison is shown in~Fig.~\ref{fig:dlg}b, and the agreement between
the analytic and the numerical solutions is excellent.

Our simplified model determines the behavior of~$b$ and~$n$, but it contains one unknown parameter, the height of the colony~$H$. In the full model, $H$~must be a function of~$D$, which is absent in the simplified model. Naturally, we expect that~$H$ is large at small~$D$, when the colony barely moves, and the biomass must accumulate in the vertical direction. In the opposite limit of~$D\to D_c$, we expect~$H\to1$, since the solution should approach the traveling front limit, for which it is easy to show that~$H=n_0$. In the Supplemental Material~\cite{si}, we derive an approximate expression for~$H(D)$, which is given by

\begin{equation}
H = \frac{1}{\sqrt{D}}.
\label{H_prediction}
\end{equation}

\noindent Figure~\ref{fig:dlg} confirms that this prediction captures
the qualitative behavior of~$\varkappa(D)$ and~$H(D)$ extremely well.
In dimensional units,~$H=\sqrt{D_n/D_b}$. Thus, the height of the
    colony scales with~$n_0$ for~$D>D_c$, but, in the square-root
    regime, it is independent of the nutrient concentration. Instead, it
    is controlled by~$D_b$, with greater motility leading to thinner
colonies.

The analysis of the square-root growth is now complete because we can obtain~$\varkappa(D)$ by combining Eqs.~\eqref{height} and~\eqref{H_prediction}; see Supplemental Material for a detailed calculation~\cite{si}. When~$D$ is close to~$D_c$, we find that~$\varkappa\approx\sqrt{D_nd/(2(1-\sqrt{D_bn_0^2/D_n}))}$ in the dimensional units. Thus,~$\varkappa$ diverges prior to the transition to traveling fronts, and the expansions are slightly faster in~$d=2$ compared to~$d=1$. For small~$D$, the effect of expansion geometry is more dramatic.

For~$d=1$, we find that~$\varkappa=n_0\sqrt{D_b/\pi}$; i.e., the rate of colony expansion is independent of~$D_n$ even though the growth is diffusion limited. The total biomass in the colony,~$B(t) = 2n_0\sqrt{D_nt/\pi}$, which equals the amount of nutrients absorbed by a stationary colony during time~$t$. Therefore, small~$D_b$ speeds up colony expansion, but does not lead to greater biomass accumulation because~$D_b$ also reduces~$H$. We then expect no selective advantage of larger~$D_b$ when the biomass motility is weak. In contrast, our results above show that both~$\varkappa$ and~$B(t)$ dramatically increase with~$D_b$ near the transition to traveling fronts, which indicates that the selective pressure on~$D_b$ could vary substantially across different growth environments. 

For~$d=2$, we find that expansions proceed faster than one dimension, and~$\varkappa = D_n^{1/4}D_b^{1/4}n_0^{1/2}/\sqrt{\ln(D_bn_0^2/D_n)/2}$, so both~$D_n$ and~$D_b$ affect the rate of colony growth. In contrast, the accumulation of the biomass depends on~$D_b$ only logarithmically:~$B(t) = -16\pi n_0D_n t/\ln(D_bn_0^2/D_n)$. Experimentally, the easiest quantity to vary is~$n_0$, and our results predict that~$\varkappa\propto n_0$ in narrow channels~($d=1$) while, up to logarithmic corrections,~$\varkappa\propto \sqrt{n_0}$ for circular colonies~($d=2$). Thus, the growth geometry controls not only the rate of colony growth, but also its dependence on the nutrient concentration.


In summary, we modified the standard model of colony growth~Eq.~\eqref{mvs_model} to make it consistent with the experiments reporting no biomass motion in regions without growth. We achieved this by making the rate of biomass redistribution proportional to the nutrient concentration, which effectively accounts for the active mechanical stresses within the growing colony. Unlike most reaction-diffusion models, Eq.~\eqref{linear_model} predicts two regimes of colony expansion: one with the radius of the colony increasing linearly with time and one with the square-root increase. The linear regime is described by the standard framework of traveling fronts, and we determined how the velocity of the expansion depends on model parameters. The square-root regime is different. The shape of the biomass profile remains invariant in time, just as in the traveling fronts regime, but the nutrient profile becomes progressively wider and more depleted. 

At first sight, it is not surprising that growth dynamics limited by nutrient diffusion lead to the~$t^{1/2}$ scaling of the colony radius with time~\cite{berg:rwib,farrell:mechanical_colonies}. We note, however, that the rate of this square-root growth could be much higher than that of a stationary colony passively absorbing nutrients. This greater expansion rate results from the nontrivial coupling between the spatial advance of the colony and the rate of nutrient acquisition summarized by Eq.~\eqref{height}.

Our model and its predictions are relatively straightforward to test experimentally. To characterize colony growth, we introduced only a handful of parameters: the initial nutrient concentration, the nutrient consumption rate, the nutrient diffusion constant, and the biomass motility. All but the last parameter can be easily measured experimentally. Thus, we have only one parameter to fit all possible observations, which include the dependence of the radius on time, colony thickness, nutrient profiles, and transitions between different growth regimes as a function of~$D$ and expansion geometry~(narrow channels vs. a Petri dish). In fact, studying colony thickness and growth dynamics as a function of~$n_0$ in different geometries appears to be the most straightforward way to validate our results.  

In closing, let us speculate that typical microbial colonies could be near the transition between the linear and the square-root growth regimes, which would explain why certain experiments report the linear growth of the radius while others report the linear increase of the area of microbial colonies without a major shift in the colony shape. In the future, we hope to explore whether the square root regime has any marked effects on the ecological and evolutionary dynamics within the colonies.

\begin{acknowledgments}
This work was supported by an NIH/NIGMS grant 1R01GM138530-01 to KSK. KJS was supported by an REU program at BU supported by NSF grant PHY-2244795. Simulations were carried out on Shared Computing Cluster at BU.
\end{acknowledgments}

\bibliographystyle{unsrtnat}
\bibliography{references}

\clearpage
\onecolumngrid
\appendix
\section*{Supplementary Material}
\renewcommand{\thefigure}{S\arabic{figure}}
\setcounter{figure}{0}
\renewcommand{\theequation}{S\arabic{equation}}
\setcounter{equation}{0}

\definecolor{soft_blue}{rgb}{0,0,0}
\newcommand{\hclr}[1]{\textcolor{soft_blue}{#1}}
\newcommand{\erf}{\mathop{\mathrm{erf}}\nolimits}
\newcommand{\erfc}{\mathop{\mathrm{erfc}}\nolimits}

This Supplemental Material contains detailed derivations and supporting data for the results reported in the main text. First, we show how the reaction-diffusion equations presented in the main text can be derived from a more detailed description of mechanical interactions between the cells, as well as between the cells and the substrate. We also discuss possible generalizations and alternative forms of the models described in the main text. The next two sections contain the analysis of the traveling fronts and square-root solutions of Eq.~\eqref{linear_model} in the main text. The fourth section provides details of the numerical methods, and the final section contains figures that provide additional details for our main results.

\section{Mathematical models}

\vspace{5pt}
\textit{Mechanics of colony growth}\\
The motion of cells within the colony can be described by a velocity field~$\bm{v}$. The biomass dynamics is then governed by

\begin{equation}
\frac{\partial b}{\partial t} = - \bm{\nabla}\cdot(b\bm{v}) + r(b,n)b,
\label{biomass_advection}
\end{equation}

\noindent which combines advection with velocity~$\bm{v}$ and growth with the growth rate~$r(b,n)$. Assuming that the mechanical properties of the colony are isotropic, the velocity field can be determined from the pressure field~$p$ by balancing the elastic and frictional forces:

\begin{equation}
-\bm{\nabla} p - \zeta_0(b,n) - \zeta_1(b,n)\bm{v} = 0,
\label{friction}
\end{equation}

\noindent where~$\zeta_0$ accounts for static friction and~$\zeta_1$ for
viscous friction. The dependence of viscosity on~$v$ is neglected, and
we also neglect the contribution of static friction in the
following~\cite{farrell:mechanical_colonies}.

To close this system of equations, we need a constitutive relationship between~$p$,~$b$, and~$n$. The standard assumption that stress is linear in strain can be expressed as

\begin{equation}
\label{constitutive_relation}
p = p_0 + \kappa(b,n)b,
\end{equation} 

\noindent where~$p_0$ is a constant, and~$\kappa$ is a measure of compressibility. Next, we express~$p$ in terms of~$b$ and substitute the result into Eq.~\eqref{friction} to find~$\bm{v}$:

\begin{equation}
\label{velocity}
\bm{v} = - \frac{\bm{\nabla}(\kappa b)}{\zeta_1}.
\end{equation}

We are now in a position to obtain the final reaction-diffusion equation that involves only~$b$ and~$n$ by substituting Eq.~\eqref{velocity} into~Eq.~\eqref{biomass_advection}. The result reads

\begin{equation}
\frac{\partial b}{\partial t} =  \bm{\nabla}\cdot\left(b\frac{\bm{\nabla}(\kappa b)}{\zeta_1}\right) + r(b,n)b.
\label{effective_model}
\end{equation}

From this equation, we can obtain all the different functional forms of the biomass redistribution term discussed in the main text. If we assume that $\kappa$ depends neither on~$n$ nor on~$b$ and that~$\zeta_1$ is a linear function of~$b$, i.e., $\zeta_1=\zeta'_1b$, then we obtain the classic FKPP equation with~$D_s=\kappa/\zeta_1'$. The first assumption is a common approximation for an elastic material, and the second assumption posits that the frictional force is directly proportional to the number of cells, which is the case for swimming motility. 

The population pressure model in Eqs.~\eqref{pressure_model}
and~\eqref{mvs_model} of the main text is obtained when both~$\kappa$ and~$\zeta_1$ are assumed to be constant. The latter assumption is equivalent to saying that, for a dense collection of cells, the frictional force with the substrate does not depend on the colony height.

If we assume that~$\zeta_1$ is constant and~$\kappa=\kappa'n$, then we
obtain a model very similar to Eq.~\eqref{linear_model} in the main text:

\begin{equation} 
\label{nutrient_compressibility_model}
\frac{\partial b}{\partial t} =  \frac{\kappa'}{\zeta_1}\bm{\nabla}(
b \bm{\nabla}(n b)) + r(b,n)b.
\end{equation}

Although different in general, these equations are approximately the
same at the colony edge, which is the only region relevant to the growth
dynamics. Ahead of the edge,~$b=0$, and, behind the edge, the biomass
approaches its limiting value~$1$ or~$H$, so its dynamics is not
relevant and all of the terms in the biomass equations are negligible.
At the colony edge,~$b$ varies much more rapidly than~$n$, at least
for~$D\ll1$ (see fig.~\ref{fig:zoom_profiles}), so the gradients of~$n$ can be neglected compared to the
gradients of~$b$, i.e., we can approximately treat~$n$ as a constant.
Therefore, we expect the same qualitative behavior as in the model discussed in the main text. Note that the linear dependence of~$\kappa$ on~$n$ is an unusual assumption. Potentially, it can be justified by considering the fact the pressure within the colony reflects both static compression and a dynamic component due to cell growth, which increases with the nutrient concentration.

Finally, our exact model is obtained by assuming that~$\zeta_1=\tilde{\zeta}/n$ and~$\kappa$ is a constant. Here, the key assumption is that the activity of growth reduces the friction with the substrate. We are not aware of any evidence supporting such an assumption, and we believe that the arguments leading to~Eq.~\eqref{nutrient_compressibility_model} provide a better justification for Eq.~\eqref{linear_model} in the main text. It is also clear that the type of reasoning outlined above could lead to more general models of the following form:

\begin{equation}
\label{general_b}
\frac{\partial b}{\partial t} =  \bm{\nabla}\cdot\left(D_1b^{\beta} n \bm{\nabla} b + D_2 b^{\beta'}\bm{\nabla} n\right) + r(b,n)b.
\end{equation} 

\begin{figure}[!h]
    \centering
    \includegraphics{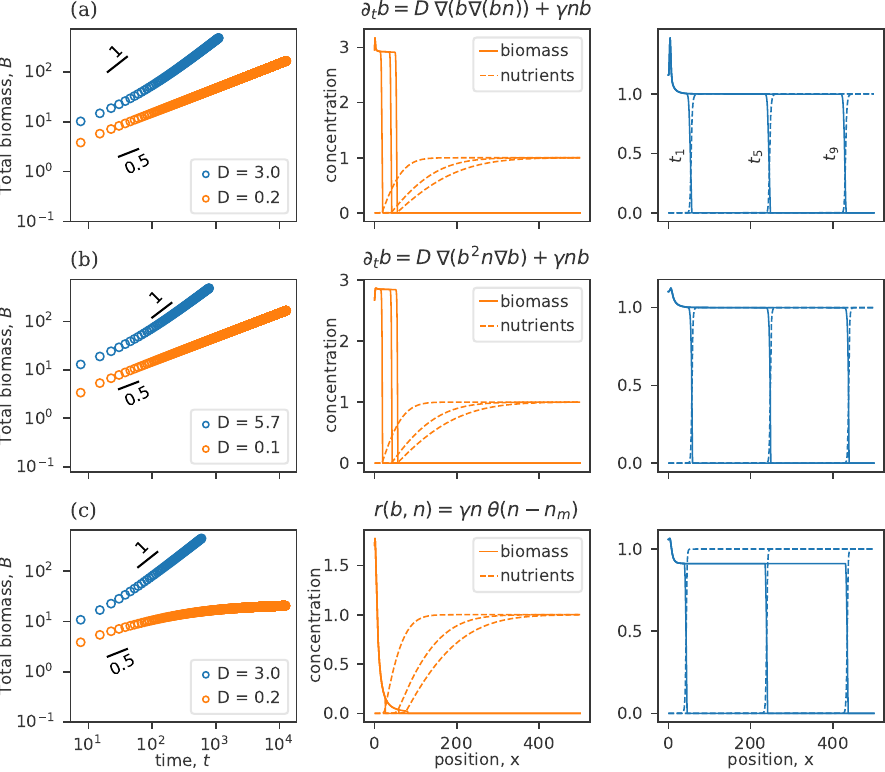}
    \caption{Transition from traveling fronts to diffusion-limited
        growth is a general feature of nutrient-dependent diffusion. For
        different models, in 1 dimension, we show the increase of biomass over time at
    low and high rates of biomass redistribution (left column), and the
corresponding biomass and nutrient profiles at 10\%, 50\%, and 90\% of the time preceding
nutrient depletion at the end of the simulation box. 
Models without a maintenance cost (rows (a) and (b)) exhibit both a linear and a square-root increase of the total biomass with time.
Introducing a maintenance cost (row (c)) drastically changes the
behavior at low dispersal. The nutrient concentration at the edge of
the colony decreases over time until it drops below the maintenance cost
threshold, at which point the biomass stops growing. For the
traveling-front growth, the biomass never reaches the maximum value set
by the initial concentration of nutrients because of the maintenance
cost.}
\label{fig:models}
\end{figure}

We briefly explored a few of such models (see Fig.~\ref{fig:models}) and found the same qualitative dynamics as reported in the main text.

\vspace{5pt}
\textit{Models of nutrient consumption}\\
In the main text, we considered a simple expression for biomass growth and nutrient consumption given by~$\gamma bn$. The actual dynamics could include a number of complications. Two most important ones are discussed below.  

First, at high nutrient concentration, both the consumption and the growth rate should saturate to a value independent of~$n$; we will denote it by~$r$ to make the connection with the FKPP equation. Then the Monod growth kinetics can be expressed as

\begin{equation}
\label{monod}
r(b,n) = \frac{\gamma n}{1+\gamma n/r},
\end{equation}

\noindent where~$\gamma$ and~$r$ are the same as in the main text. In particular,~$r$ reflects the maximum growth rate~($\mu_{\mathrm{max}}$ in the Monod equation) and the half-saturation constant is~$r/\gamma$. 

The nutrient consumption could be different from~$r(b,n)$ if the yield depends on the nutrient concentration. A simple Michaelis-Menten model would lead to

\begin{equation}
\label{michaelis_menten}
\frac{\partial n}{\partial t} = D_n\bm{\nabla}^2n - \frac{\gamma nb}{1+\gamma n/r'},
\end{equation}

\noindent with~$r'\ge r$. For large~$D$, we still expect traveling fronts with velocities that scale as~$\sqrt{D_b}$ when~$D_b\to+\infty$ because~$D_n$ is irrelevant and~$D_b$ is the only parameter containing the dimension of length. The dependence of~$v$ on~$n_0$ will be different depending on the values of~$\gamma$ and~$r$. For small~$D$, the nutrient concentration at the edge of the colony decreases with time and eventually falls well below the half-saturation constant. At this point, the more complex models based on the Monod and Michaelis-Menten kinetics reduce to the simpler terms studied in the main text. Therefore, we expect no major differences for small~$D_b$ near or below the transition to square-root growth.

Second, at very low nutrient concentrations, one has to account for the metabolic cost of cell maintenance because it becomes a significant fraction of total energy expenditure when the growth rates are low. A simple, if a little crude, way to account for these effects is to introduce a minimal nutrient concentration~$n_m$ required for growth:

\begin{align}
 &r(b,n) = \frac{\gamma n}{1+\gamma n/r}\theta(n-n_m)\\
 &\frac{\partial n}{\partial t} = D_n\bm{\nabla}^2n  - \frac{\gamma nb}{1+\gamma n/r}.
\label{maintenance}
\end{align}

This modification makes no appreciable effect on the dynamics of traveling fronts except in the immediate vicinity of the transition
when~$v$ is small. Indeed, we expect that~$n_e\approx1-1/D$; see Eq.~\eqref{traveling_front_solution}, so~$n_e$ stays above~$n_c$ and the maintenance cost has no effect on the growth term at the front. Behind the front, the maintenance cost reduces the biomass concentration in the colony bulk below~$n_0$. 

The maintenance cost, however, completely destroys
the asymptotic~$t^{1/2}$ growth because the colony stops growing
altogether once the nutrient concentration falls below~$n_c$ at the
colony edge. Since~$n_e\propto t^{-1/2}$, this growth arrest is inevitable. That said, the time required to deplete the nutrients below~$n_c$ could be quite
long, and the intermediate behavior is given by the square-root growth
as before. In other words, the~$t^{1/2}$ behavior for growth becomes an
intermediate asymptotic. This behavior is illustrated in
Fig.~\ref{fig:models}, in the limit of large $r$.

\section{Traveling-front solutions}

By definition, traveling wave solutions are plane waves, i.e.~$b(t,\bm{r})=b(x-vt)$ for waves traveling in the positive $x$-direction. When growth instabilities are suppressed by internal dynamics, surface tension, or other mechanisms, the planar traveling fronts are also good approximations for growing circular colonies. Indeed, the width of traveling fronts~(and the nutrient depletion layer) are finite and, as we show below, are typically small; therefore, the effect of circular geometry should decrease as the inverse of the colony radius. For the FKPP equation, the effects of front curvature are discussed in Ref.~\cite{murray:book}. We limit our discussion of traveling front solution to planar wave fronts, which are effectively one-dimensional.

\vspace{5pt}
\textit{Properties of the solutions and the shooting method}\\
In this section, we analyze the behavior of the traveling front
solutions of Eq.~\eqref{linear_model} near the colony edge and deep in
the colony bulk. This analysis provides the initial and final conditions
for the shooting method to determine the front profiles and velocities
numerically. We also derive the analytical approximation for~$v(D)$
given by~Eq.~\eqref{v_analytic} by matching the initial and final conditions. 

We begin by stating the reaction-diffusion model in the nondimensionalized variables:

\begin{equation}
\begin{aligned}
& \frac{\partial b}{\partial t} = D\frac{\partial }{\partial x}\left(bn\frac{\partial b}{\partial x}\right) +  bn,\\
& \frac{\partial n}{\partial t} = \frac{\partial^2 n}{\partial x^2} -  bn .\\
\end{aligned}
\label{pde}
\end{equation} 

\noindent with the initial nutrient concentration equal to~$1$. We then substitute the traveling wave ansatz into this partial differential equation. Specifically, we assume that~$b$ and~$n$ are functions of~$z$ only, with~$z=x-vt$. The resulting system of ordinary differential equations reads

\begin{equation}
\begin{aligned}
& D(bnb')' + vb' + bn = 0 ,\\
& n'' + vn' -  bn = 0 ,\\
\end{aligned}
\label{ode}
\end{equation} 

\noindent where primes denote derivatives with respect to~$z$. To solve these equations, we also need the boundary conditions:

\begin{equation}
\begin{aligned}
& b(t,-\infty) = 1,\\
& b(t,+\infty) = 0,\\
& n(t,-\infty) = 0,\\
& n(t,+\infty) = 1.\\
\end{aligned}
\end{equation}

\noindent Here, only the first equation may require an explanation. For a traveling front, the amount of nutrient consumed per unit of time in the entire system is~$vn_0=v$ since~$n_0=1$ after nondimensionalization. Indeed,~$\frac{d}{dt}\int ndx = -v\int n'dx = -v(1-0)$.  Since all of this nutrient is converted into biomass, the rate of biomass production across the entire system must also equal~$v$. Thus,~$v=\frac{d}{dt}\int bdx = -v\int b'dx = -v(0-b(t,-\infty))$ or equivalently~$b(t,-\infty) = 1$.

Since the nutrient concentration at the colony edge is finite, we expect
the traveling front to have the same asymptotic behavior at the colony
edge as the models described by Eqs.~\eqref{pressure_model}
and~\eqref{mvs_model} of the main text. Specifically, we anticipate that~$b(z)=0$ for~$z\ge0$ assuming that~$z=0$ at the colony edge~\cite{muller:Db_model,kawasaki:exact_dispersal,birzu:dispersal}. Therefore, there are three important regions to analyze:~$z>0$, negative~$z$ near the colony edge, and negative~$z$ far away from the edge.

In the rightmost region with~$b=0$, the biomass equation is satisfied for any value of~$n$, and the nutrient equation becomes independent of~$b$ and is easily solved. The nutrient profile is given by

\begin{equation}
\label{n_ahead}
n(z) = 1 - (1-n_e)e^{-vz},
\end{equation}

\noindent where~$n_e=n(0)$ is the nutrient concentration at the edge of the colony that needs to be determined.

Within the colony, the nutrient concentration is still given by Eq.~\eqref{n_ahead} in the immediate vicinity of~$z=0$ because~$n(z)$ is continuously differentiable. To determine the behavior of~$b(z)$, we use the ansatz from~Ref.~\cite{muller:Db_model}, which is equivalent to the dominant balance of the first two terms in the equation~(i.e., we neglect the last term because it does not contain derivatives, which are large at the front). Specifically, we look for~$b(z)=-Az$, where~$A$ is a positive constant that can be determined upon the substitution of the ansatz into the equation:

\begin{equation}
DA^2(n+zn') + vA + Azn = 0.
\end{equation}

\noindent In the limit of~$z\to-0$, we find that~$A = -v/(Dn_e)$, and

\begin{equation}
\label{b_edge}
b(z) = -\frac{v}{Dn_e}z.
\end{equation}

Finally, for large negative~$z$, we first solve the equation for~$n$ by assuming that~$b\approx1$, and then solve the equation for~$b$ by assuming that~$n$, $b'$, and~$b''$ are all small; in fact, they are of the same order due to the exponential approach to the boundary conditions. The linear equation for the nutrient is solved by the standard method and the solution reads

\begin{equation}
\label{n_behind}
n = C e^{\lambda z},
\end{equation}

\noindent where~$C$ is a yet unknown constant, and~$\lambda$ is given by

\begin{equation}
\label{lambda}
\lambda = \frac{\sqrt{v^2+4}-v}{2}.
\end{equation}

The biomass equation can be written as

\begin{equation}
D(nb'^2+bb'n'+bnb'') + vb' + bn = 0,
\label{b_bulk}
\end{equation} 

\noindent and it is clear that all the terms with~$D$ are higher order than the remaining two terms. Upon further approximating~$b$ by~$1$, we find

\begin{equation}
b' = -\frac{n}{v}=-\frac{C}{v}e^{\lambda z},
\end{equation} 

\noindent which can be integrated to give the biomass profile

\begin{equation}
b = 1-\frac{C}{v\lambda}e^{\lambda z}.
\label{b_behind}
\end{equation}

This completes the analysis necessary to set up the shooting method. We integrate Eq.~\eqref{ode} starting from an arbitrary point~$z_i$ with~$n(z_i)$ and its derivative determined from Eq.~\eqref{n_behind} and~$b$ and its derivative determined from~Eq.~\eqref{b_behind}. Note that~$C$ must be chosen such that~$n(z_i)\ll1$. The integration stops when~$b=0$ or the solution becomes unphysical. At this final point, we have a numerically computed value of~$n_e$, but the solution must also satisfy Eqs.~\eqref{n_ahead}. This condition is not satisfied for an arbitrary~$v$, so we perform a search to find~$v$ that results in a feasible solution.

\vspace{5pt}
\textit{Analytical approximation}\\
We can obtain an approximate analytical solution for the traveling front profile and velocity by matching the solutions at large negative~$z$ and at~$z=0$. That is,~Eq.~\eqref{b_behind} is matched with~Eq.~\eqref{b_edge}, and~Eq.~\eqref{n_behind} is matched with~Eq.~\eqref{n_ahead}. This results in four conditions for~$b$,~$b'$,~$n$,~and~$n'$:

\begin{equation}
\begin{aligned}
	& 1-\frac{C}{v\lambda} = 0,\\
	& -\frac{C}{v} = -\frac{v}{Dn_e},\\
	& C = n_e,\\
	& C\lambda  = v(1-n_e).\\
\end{aligned}
\end{equation}

Upon eliminating~$C$ and~$n_e$, we are left with two conditions:

\begin{equation}
\begin{aligned}
	& \lambda^2 D = 1,\\
	& \lambda^2 + \lambda v - 1 = 0.\\
\end{aligned}
\end{equation}

The second condition is exactly the same condition as Eq.~\eqref{lambda}, so we have a unique solution for~$v(D)$ and other parameters:

\begin{equation}
\begin{aligned}
	&  v = \sqrt{D} - \frac{1}{\sqrt{D}},\\
	& n_e = 1-\frac{1}{D},\\
	& C = 1-\frac{1}{D},\\
	& \lambda = \frac{1}{\sqrt{D}}.\\
\end{aligned}
\label{traveling_front_solution}
\end{equation}

The expression for~$v$ is the results stated in Eq.~\eqref{v_analytic} in the main text. Note that, if we restore the dimensional units, we need to multiply~$v$ by~$\sqrt{D_n\gamma n_0}$. In the limit of large~$D$, this gives~$v=\sqrt{D_b\gamma n_0^3}$, which means that the nutrient diffusion constant does not affect the velocity in the large~$v$ limit.

\section{Diffusion-limited growth}

Numerical simulations show that, for~$D<D_c$, the biomass profile approaches a step function moving outward as~$t^{1/2}$. Moreover, the nutrient concentration at the edge of the colony drops to nearly zero, suggesting that the nutrients are immediately consumed when they reach the colony; see~Fig.~\ref{fig:nutrient_decay}. In this section, we analyze this limit analytically. 

Unlike for the traveling front solutions, the growth dynamics depends on the geometry of the problem even in the long time limit. Indeed, both the spatial extent of nutrient depletion and the radius of the colony grow as~$t^{1/2}$, so their ratio stays constant and the growth dynamics do not reduce to the planar geometry. In the following, we assume that the growth instability is suppressed, e.g. by strong surface tension, and analyze spherically symmetric growth in~$d$ spatial dimensions. Thus, $d=1$~corresponds to planar fronts, which could be realized in narrow channels, $d=2$~corresponds to regular colonies grown on the surface of a Petri dish, and $d=3$~corresponds to spherical aggregates akin early stage tumors.  

We start with the following set of equations, which describe the asymptotic dynamics in the diffusion-limited regime:

\begin{equation}
\begin{aligned}
& \frac{\partial n}{\partial t} = \frac{\partial^2 n}{\partial r^2} + \frac{d-1}{r}\frac{\partial n}{\partial r}, \\
& H\frac{dr_e}{dt} = \left.\frac{\partial n}{\partial r}\right|_{r=r_e}, \\
& n(t,r_e) = 0,\\
& n(t,+\infty) = 1,\\
\end{aligned}
\label{diffusion_problem}
\end{equation}

\noindent where we used~$r$ to denote the distance from the colony center in any number of dimensions. When we are explicitly considering an one-dimensional situation, we revert to using~$x$ as the spatial coordinate. The subscript~$e$ refers to the colony edge.

Given that these equations describe diffusive dynamics, it is natural to seek the solution in the following form:

\begin{equation}
\begin{aligned}
& r_e(t) = 2\varkappa \sqrt{t}, \\
& n(t,r) = n(\zeta),\\
& \zeta = \frac{r}{2\sqrt{t}},\\
& n(\zeta=\varkappa) = 0,\\
& n(\zeta=+\infty) = 1,\\
\end{aligned}
\end{equation}

\noindent where the fourth equation follows from~$n(t,r_e)=0$ and the ansatz for~$r_e(t)$. 

Upon replacing~$n(t,r)$ by~$n(\zeta)$, we find

\begin{equation}
\frac{d^2n}{d\zeta^2} + \left(2\zeta +\frac{d-1}{\zeta}\right) \frac{dn}{d\zeta} = 0,
\end{equation}

\noindent which can be easily solved:

\begin{equation}
n = \frac{\int_{\varkappa}^{\zeta}dp p^{1-d}e^{-p^2}}{\int_{\varkappa}^{+\infty}dp p^{1-d}e^{-p^2}}= 1 - \frac{K_d(\zeta)}{K_d(\varkappa)},
\label{n_solution}
\end{equation} 

\noindent where

\begin{equation}
K_d(y) = \int_{y}^{+\infty}dp p^{1-d}e^{-p^2}.
\end{equation}

\noindent Note that~$K_d$ is related to the upper incomplete gamma function, but we do not find that this connection is worth exploiting and instead specifically state~$K_d$ for one, two, and three dimensions below:

\begin{equation}
\begin{aligned}
& K_1(y) = \frac{\sqrt{\pi}}{2}\erfc(y) ,\\
& K_2(y) = \frac{1}{2}\mathrm{E_1}(y^2) = \frac{1}{2} \int_{y^2}^{+\infty}\frac{e^{-q}}{q}dq, \\
& K_3(y) = \frac{e^{-y^2}}{y} -\sqrt{\pi}\erfc(y), \\ 
\end{aligned}
\label{Kd}
\end{equation}

\noindent where~$\erfc(y)$ and~$\mathrm{E_1(y)}$ are the complimentary error function and exponential integral respectively.

From Eq.~\eqref{n_solution}, we can determine the nutrient flux at~$r_e$ and thus determine~$\frac{dr_e}{dt}$:

\begin{equation}
H\frac{dr_e}{dt} = \left.\frac{dn}{d\zeta}\right|_{\zeta=\varkappa}\frac{\partial\zeta}{\partial r}.
\end{equation}

We can also determine~$\frac{dr_e}{dt}$ by differentiating~$r_e=2\varkappa\sqrt{t}$ and thus obtain the following self-consistency condition:

\begin{equation}
H^{-1} = 2\varkappa^d e^{\varkappa^2}\int_{\varkappa}^{+\infty}p^{1-d}e^{-p^2}dp = 2\varkappa^d e^{\varkappa^2} K_d(\varkappa) ,
\label{H_varkappa}
\end{equation}

\noindent which is Eq.~\eqref{height} in the main text. This completes the solution of~Eq.~\eqref{diffusion_problem} because we have determined the functional forms of~$x_e(t)$ and~$n(t,x)$ in terms of~$\varkappa$, and found an implicit equation for~$\varkappa$ in terms of~$H$, which is a parameter in the simplified model. 

The implicit equation for~$\varkappa(H)$ can be easily analyzed in the limit of small and large~$\varkappa$ by expanding~$K_d(y)$ for small and large values of its argument. For slow expansions with~$\varkappa\ll1$, we find

\begin{equation}
\begin{aligned}
& H^{-1} = \sqrt{\pi}\varkappa,\;\; d=1,\\
& H^{-1} = -2\varkappa^2\ln\varkappa, \;\; d=2,\\
& H^{-1} = 2\varkappa^2, \;\; d=3,\\
& H^{-1} = \frac{2\varkappa^2}{d-2},\;\; d>2,\\
\end{aligned}
\label{small_varkappa_H}
\end{equation}

\noindent which can be inverted to find~$\varkappa(H)$:

\begin{equation}
\begin{aligned}
& \varkappa = \frac{1}{\sqrt{\pi}H} = \sqrt{\frac{D}{\pi}},\;\; d=1,\\
& \varkappa = \frac{1}{\sqrt{H\ln(H)}} \approx \frac{D^{1/4}}{\sqrt{\ln(D)/2}}, \;\; d=2,\\
& \varkappa = \frac{1}{\sqrt{2H}}=\frac{D^{1/4}}{\sqrt{2}}, \;\; d=3,\\
& \varkappa = \sqrt{\frac{d-2}{2H}}=D^{1/4}\sqrt{\frac{d-2}{2}}, \;\; d>2,\\
\end{aligned}
\label{small_varkappa}
\end{equation}

\noindent where we used the result that~$H=1/\sqrt{D}$, which is derived below. Note that for~$d=2$ the asymptotic expression given above might be better approximated by the exact solution of~$H^{-1} = -2\varkappa^2\ln\varkappa$, which reads~$\varkappa = \exp[\mathrm{W_{-1}}(-H^{-1})/2]$, where~$\mathrm{W_{-1}}(\cdot)$ is the Lambert~$W$ function. Note that for a given value of~$H$ or~$D$, the expansion rate~$\varkappa$ increases with~$d$, with the biggest difference between~$d=1$ and~$d=2$. This speed up reflects the increasing ratio of the surface to the volume in higher dimensions, i.e. the greater amount of nutrient that can diffuse from the outside in a given solid angle.

To restore the dimensional units, we need to multiply the dimensionless value of~$\varkappa$ by~$\sqrt{D_n}$. In one dimension, we find that~$\varkappa= n_0\sqrt{D_b/\pi}$ for small~$D$. Thus, the rate of the colony expansion is independent of~$D_n$ even though the growth is diffusion limited. In two dimensions, we find that, for small~$D$,~$\varkappa = D_n^{1/4}D_b^{1/4}n_0^{1/2}/\sqrt{\ln(D_bn_0^2/D_n)/2}$, so both~$D_n$ and~$D_b$ affect the rate of colony growth. Experimentally, the easiest quantity to vary is~$n_0$, and our results predict that~$\varkappa\propto n_0$ in narrow channels while~$\varkappa\propto \sqrt{n_0}$ for circular colonies~(up to logarithmic corrections). Thus, the growth geometry controls not only the rate of colony growth, but also its dependence on the nutrient concentration. Note that the scaling of the traveling front velocity with~$n_0$ is different from that of~$\varkappa$. For~$D\gg1$, we expect that~$v\propto n_0^{3/2}$.

In addition to the rate of square-root expansion~$\varkappa$, we can characterize the growth of the colony by the rate of biomass accumulation~$B(t) = S_dH(2\varkappa\sqrt{t})^d$, where~$S_d$ is the area of unit sphere embedded in~$d$ dimensions. Using our results above we find that

\begin{equation}
\begin{aligned}
& B(t) = \frac{2t^{1/2}}{\sqrt{\pi}} ,\;\; d=1,\\
& B(t) = \frac{8\pi t}{\ln(H)} = -16\pi t\frac{1}{\ln(D)} , \;\; d=2,\\
& B(t) = \frac{16\pi t^{3/2}}{\sqrt{2H}}= 8\sqrt{2}\pi t^{3/2}D^{1/4}, \;\; d=3,\\
& B(t) = S_d2^dt^{d/2}H^{\frac{2-d}{2}}\left(\frac{d-2}{2}\right)^{d/2}=S_d2^dt^{d/2}D^{\frac{d-2}{4}}\left(\frac{d-2}{2}\right)^{d/2}, \;\; d>2.\\
\end{aligned}
\label{small_varkappa_B}
\end{equation}

For~$d=1$, we can set~$D=0$ and observe that our expression for $B(t)$ matches the expected amount of nutrient absorbed by a stationary colony.~(In dimensional units,~$B(t) = 2n_0\sqrt{D_nt/\pi}$; note,~$H$ scales as~$n_0$ not as distance.) Therefore, a small increase in~$D$ above zero produces only moderate increase in~$B(t)$ compared to a stationary colony. For large~$\varkappa$, however, a moving colony consumes much more nutrient, and higher biomass motility~$D$ is greatly beneficial for any~$d$; see Eq.~\eqref{large_varkappa}.

For~$d=2$,~$B(t) = -16\pi n_0D_n t/\ln(D_bn_0^2/D_n)$ in dimensional units. The predicted logarithmic dependence on~$D_b$, however, could be influenced by long transients due to the competition between the solution with small, but nonzero~$D_b$ and the solution for a stationary colony of a nonzero radius, which also leads to~$B(t)\propto t$. In fact, large inoculations, could produce colonies for which the latter solution dominates. Small inoculations, e.g. started from a few cells, should however be described by our solution provided~$D_b$ is not too small. 

For $d=3$, our results may not be directly applicable to actual growing populations, at least in the limit of~$H\to+\infty$, because~$H$ corresponds to the amount of biomass packed inside physical space. For~$d=1$ and~$d=2$, the biomass can escape in the third dimension, but this is not possible for~$d=3$, where~$H$ corresponds to biomass compression. Given that cell are nearly incompressible, we do not expect that three dimensional aggregates can achieve large values of~$H$, and their motility should be described by the pressure-driven model of an incompressible fluid.

For fast expansions with~$\varkappa\gg1$, which occur for~$H\to1$ and~$D\to1$, Eq.~(\ref{H_varkappa}) yields the following asymptotic results:

\begin{equation}
 H^{-1} = 1 - \frac{d}{2\varkappa^2},
\label{large_varkappa_H}
\end{equation}

\noindent which can be inverted to find~$\varkappa(H)$:

\begin{equation}
\varkappa = \sqrt{\frac{Hd}{2(H-1)}} = \sqrt{\frac{d}{2(1-\sqrt{D})}},
\label{large_varkappa}
\end{equation}

\noindent where we again used the result that~$H=1/\sqrt{D}$. As for~$\varkappa\ll1$, we find that, for~$\varkappa\gg1$, colonies expand faster in higher dimensions, but the speed up from~$d=1$ to~$d=2$ is less dramatic. More importantly, we find that the rate of the square-root expansion diverges as~$H$ approaches unity from above. These results are easy to understand: The bigger the~$H$, the more nutrient is necessary to advance the colony edge forward. Therefore, thick colonies put most of their biomass growth in the vertical direction and expand slowly, while thin colonies put all of their growth into the horizontal direction and expand faster. This feedback between local biomass accumulation and outward expansion is nonlinear because the faster the outward expansion, the greater the amount of nutrient that the colony consumes. Equation~\eqref{H_varkappa} captures this nonlinear feedback quantitatively.

\vspace{5pt}
\textit{Derivation of Eq.~\eqref{H_prediction}}\\
Figures~1 and~3 in the main text provide strong evidence that the simplified model defined by Eq.~\eqref{diffusion_problem} indeed describes the long-time behavior of Eq.~\eqref{pde}. The only missing link in this correspondence is the dependence of~$H$ on~$D$, which are the sole parameters in the respective models. Intuitively, we expect that larger~$D$ should result in greater~$\varkappa$ and, therefore, lower~$H$. Here, we provide an approximate derivation of~$H(D)$, which is stated as Eq.~\eqref{H_prediction} in the main text. 

As we mentioned above, the biomass profile remains sharp, i.e, it has a constant width, even though the region of nutrient depletion grows as~$t^{1/2}$. Therefore, we can neglect the curvature of the biomass front. We show below that the nurient equation also reduces to the plane-wave geometry, and, therefore, Eq.~\eqref{H_prediction} holds for any~$d$. Our approach is to follow the steps of the previous section on the traveling-front solutions, with two important differences. First, the biomass density behind the front now saturates at~$H$ instead of~$1$. Second, we now treat~$n_e=n(t,r_e(t))$ and~$v=\frac{dr_e}{dt}=\frac{\varkappa}{\sqrt{t}}$ as slowly varying functions of time. That is, we do not neglect their time dependence, but we do neglect their time derivatives because they vanish much faster, as~$t^{-3/2}$, at long times.

We start with the nutrient profile, which we assume is of the form~$n_e(t)\tilde{n}(z)$, where~$z=r-r_e(t)$. For~$r>r_e$,~Eq.~\eqref{n_solution} provides a good approximation to~$n(t,r)$. However, since by assumption~$n_e(t)=0$, we cannot use this equation directly. Instead, we compute the flux of the nutrient because it does not vanish and therefore could be used to link the solutions for~$r>r_e$ and~$r<r_e$. This flux is given by~$H\frac{dr_e}{dt}= \frac{H\varkappa}{\sqrt{t}}$, so we require that

\begin{equation}
\label{ne_bc}
\left.\frac{\partial n}{\partial r}\right|_{r=r_e} = \frac{H\varkappa}{\sqrt{t}}.
\end{equation}

Inside the colony, the nutrient profile approximately satisfies~Eq.~\eqref{ode} with the modifications specified above:

\begin{equation}
\frac{d^2\tilde{n}}{dz^2} + \frac{d-1}{r}\frac{d\tilde{n}}{dz}+v(t)\frac{d\tilde{n}}{dz}-H\tilde{n}=0.
\end{equation}

Note that both~$v(t)$ and~$r_e^{-1}$ decrease as~$t^{-1/2}$, and a nontrivial limit exists even when~$v=0$ and~$r\to+\infty$:

\begin{equation}
\frac{d^2\tilde{n}}{dz^2} - H\tilde{n}=0.
\end{equation}

Thus, to the leading order, we find that~$\tilde{n} = e^{\sqrt{H}z}$, and, therefore,

\begin{equation}
n(t,r) = n_e(t)e^{\sqrt{H}(r-r_e(t))}.
\label{n_inside}
\end{equation}

Upon combining~Eqs.~\eqref{ne_bc} and~\eqref{n_inside}, we obtain the nutrient concentration at the colony edge:

\begin{equation}
\label{ne_diffusion}
n_e(t) = \frac{\sqrt{H}\varkappa}{\sqrt{t}}.
\end{equation}

Next, we turn to the biomass profile,~$b(t,r)=b(r-r_e(t))$. Near the colony edge, all the steps leading to~Eq.~\eqref{b_edge} remain valid, and we immediately obtain that

\begin{equation}
\label{b_edge_diffusion}
b(z) = -\frac{v(t)}{Dn_e(t)}z = -\frac{1}{D\sqrt{H}}z.
\end{equation}

\noindent Note that the time dependence of~$v$ and~$n_e$ cancels out, so the shape of~$b(z)$ has no time dependence, in agreement with simulations.

The final equation that we need is~Eq.~\eqref{b_bulk}, which we use to obtain the biomass profile for large negative~$z$. This equation also remains unchanged given our assumption of slowly varying~$v(t)$ and large~$r$. As before, we neglect the first three terms that contain two spatial derivatives because they are much smaller than the remaining terms, given than both~$n$ and~$b$ approach their limiting values exponentially in~$z$. We then conclude that

\begin{equation}
\label{b_behind_diffusion}
b' = -\frac{bn}{v} = -\frac{Hn_e(t)\tilde{n}(z)}{v(t)}= -H^{3/2}\tilde{n}(z),
\end{equation}  

\noindent where the time dependence again cancels out.

Similar to our approximation for traveling fronts, we now match the behavior of~$b$ for large and small~$z$, i.e., we match~Eqs.~\eqref{b_edge_diffusion} and~\eqref{b_behind_diffusion}. This gives

\begin{equation}
\frac{1}{D\sqrt{H}}=H^{3/2},
\end{equation}

\noindent or equivalently

\begin{equation}
H=\frac{1}{\sqrt{D}},
\label{H}
\end{equation}

\noindent which is the same as~Eq.~\eqref{H_prediction} in the main text. The general dependence of~$\varkappa$ on~$D$ follows from~Eqs.~\eqref{H} and~\eqref{H_varkappa} as discussed above.

Note that, if we restore the dimensional units of~$H$ then we need to multiply it by~$n_0$ since~$H$ is a measure of biomass not distance. Hence,~$H=\sqrt{D_n/D_b}$, i.e., it does not depend on~$n_0$ and~$\gamma$.

\section{Simulations}

Equation~\eqref{pde} and other models were solved using a finite
difference method. The biomass equation was solved using an explicit
method, and the nutrient equation was solved using the Crank-Nicolson
method~\cite{numerical:recipes}. 
All simulations were started from a small initial biomass concentration
at the left edge of the simulation box, with uniform nutrient
concentration, and were ran until the nutrients at the right edge of the
simulation box started to be consumed. 
All the figures were obtained with a no-flux boundary condition for the
biomass and a Dirichlet boundary condition for the nutrients. 
We also performed simulations with a no-flux boundary condition for the
nutrients and the result are identical, since we stop the simulation
as soon as boundary effects materialize.
The simulation and analysis codes are available at
\url{https://github.com/lbrezin/fronts-to-diffusion}.

\section{Supporting figures}

\begin{figure}[!h]
    \centering
    \includegraphics{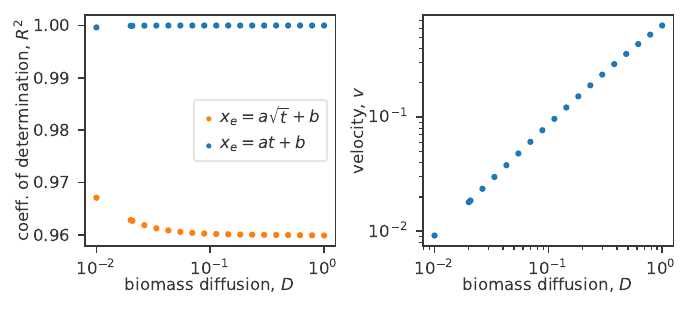}
    \caption{The model introduced in Eq.~\eqref{mvs_model} in the main text, with the
        motility independent of nutrient concentration, does not exhibit the transition
seen in our model. There is a traveling wave solution at low $D$ with a
velocity that scales linearly with $D$ for $D\ll1$
\cite{muller:Db_model}. The uptick seen at low $D$ for the square-root
fitting is due to rapidly increasing transient times as $D \to 0$.}
    \label{fig:nutrient_independent}
\end{figure}

\begin{figure}[!h]
    \centering
    \includegraphics{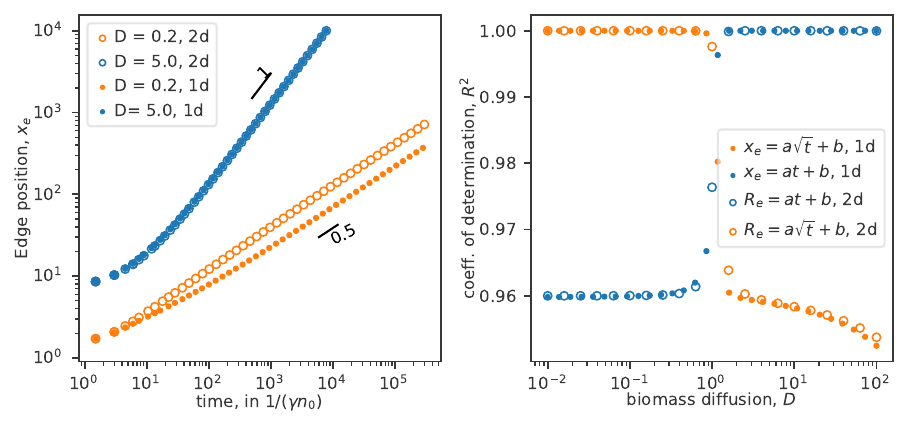}
    \caption{Comparing simulations of Eq.~\eqref{linear_model} in 1 and 2 dimensions. The
    traveling wave solution is identical in both cases, and the
transients are different in the diffusion-limited growth, but the
transition remains the same at $D_c = 1$.}
    \label{fig:transition_12d}
\end{figure}

\begin{figure}[!h]
    \centering
    \includegraphics{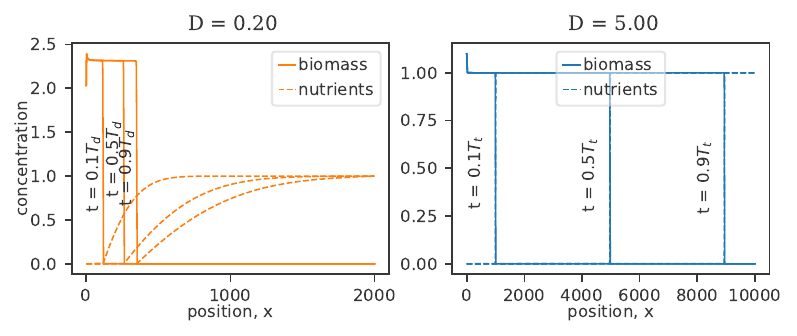}
    \caption{Biomass and nutrient profiles in 1d for diffusion-limited
    growth and traveling front. The profiles are similar to the 2d case shown
in the main text in Fig.~\ref{fig:transition}c-d.}
    \label{fig:profiles_1d}
\end{figure}

\begin{figure}[!h]
    \centering
    \includegraphics{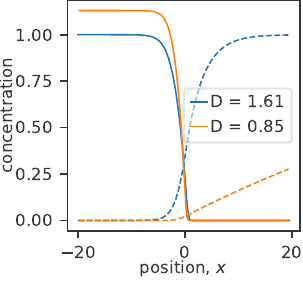}
    \caption{Close-up view of the biomass and nutrient profiles near the
    edge of the colony, for a 1d expansion. For the traveling wave
    solution at $D>D_C$ (blue lines) the size of the depletion layer is
    comparable to the width of the biomass front. For the
    diffusion-limited growth solution at $D<D_C$ (yellow lines), the
    size of the depletion layer is much greater than the width of the
    biomass front}
    \label{fig:zoom_profiles}
\end{figure}

\begin{figure}[!h]
    \centering
    \includegraphics{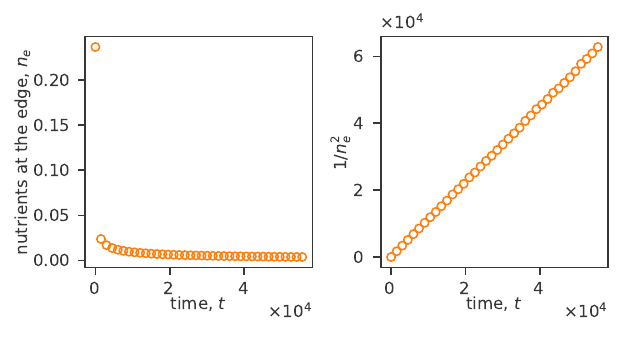}
\caption{The nutrient concentration at the colony edge vanishes an~$t^{-1/2}$ as predicted by~\protect{\eqref{ne_diffusion}}.}
\label{fig:nutrient_decay}
\end{figure}

\begin{figure}[!h]
    \centering
    \includegraphics{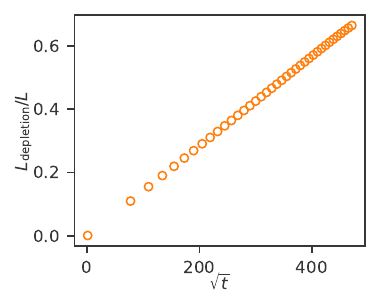}
    \caption{The size of the depletion layer $L_\text{depletion}$ increases as $t^{1/2}$. We
 see a linear relationship between $L_\text{depletion}$ normalized
        by the size of the system as a function of the square root of
        time, for the
        diffusion-limited growth. We define
$L_\text{depletion}$ as the distance between the leading edge where the
biomass is 5\% of its maximal value and the first point where there is
full nutrient availability (nutrients above 99\% of their initial value). 
}
    \label{fig:depletion_layer}
\end{figure}

\begin{figure}[!h]
    \centering
    \includegraphics{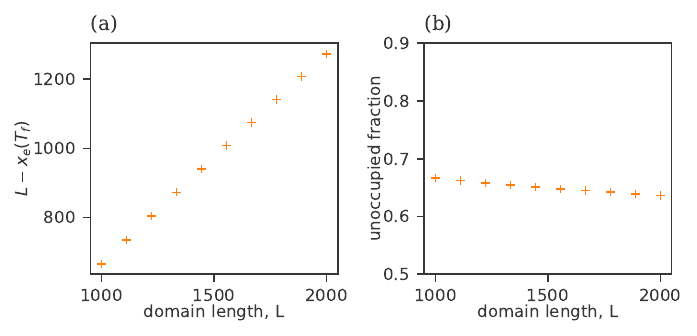}
    \caption{The distance between the edge of the system $L$ and the edge of the
        colony once all nutrients are depleted $x_e(T_f)$ is expected to increase as a function of the system size $L$ in the
    diffusion-limited regime (a), while we expect the unoccupied
    fraction $\frac{L-x_e}{L}$ to stay roughly the same (b).
    Increasing system size is equivalent to increasing total simulation
    time, therefore increasing the size of the depletion layer until it
    reaches the boundary.
    For the fraction, both the colony edge position and the depletion layer
    increase as $t^{1/2}$, such that their ratio is expected to be
    independent of time. Therefore, we expect the unoccupied fraction to 
    be constant.} 
    \label{fig:distance_edge}
\end{figure}

\end{document}